\documentstyle[multicol,aps,epsfig]{revtex}
\newcommand{\bm}[1]{\mbox{\boldmath $#1$}}
\begin{document}
\draft
\input epsf 

\title{Intermittency in two-dimensional Ekman-Navier-Stokes turbulence}

\author{G. Boffetta$^{1,2}$, A. Celani$^{3}$, 
S. Musacchio$^{1,2}$, M. Vergassola$^{4}$}
\address{$^{1}$Dipartimento di Fisica Generale, Universit\`a di Torino,
         Via Pietro Giuria 1, 10125 Torino, Italy.}
\address{$^{2}$INFM Sezione di Torino Universit\`a, Corso Raffaello 30,
10125 Torino, Italy.}
\address{$^{3}$CNRS, INLN, 1361 Route des Lucioles, 06560 Valbonne, 
France.} 
\address{$^{4}$CNRS, Observatoire de la C\^ote d'Azur, B.P. 4229, 06304 
Nice Cedex 4, France.} 
\date{\today}

\maketitle

\begin{abstract}
We study the statistics of the vorticity field in two-dimensional 
Navier-Stokes turbulence with a linear Ekman friction.  
We show that the small-scale vorticity fluctuations are intermittent, 
as conjectured by Nam {\it et al.\/} [{\it Phys. Rev. Lett.\/} {\bf 84} 
(2000) 5134]. The small-scale statistics of vorticity 
fluctuations coincides with the one of a passive scalar with finite lifetime
transported by the velocity field itself.
\end{abstract}

\begin{figure}[hb]
\epsfxsize=250pt
\centerline{\hspace{3cm}\epsfbox{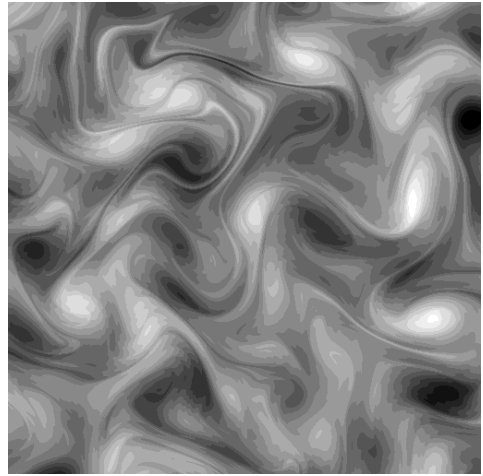}\vspace{0.2cm}}
\caption{Snapshot of the vorticity field. 
The numerical integration of Eq.~\protect(\ref{eq:1}) 
is performed by a fully dealiased 
pseudospectral code with a second-order Runge-Kutta scheme, 
on a doubly periodic square domain of size $L = 2\pi$     
at different resolutions: $N^2 = 512^2, 1024^2, 2048^2$ grid points.
(The present figure is taken from the $1024^2$ run).
Standard viscosity or hyperviscosity (depending on the resolution)
is used to remove the remnant enstrophy flux at small scales.    
The large-scale forcing $f_\omega$ is Gaussian, 
$\delta$-correlated in time, and limited to a shell of wavenumbers
around $k_f=2\pi/L$ ($k_f=2$ for $N=512,2048$, and $k_f=8$ for 
$N=1024$).  
At variance with other choices for $f_\omega$ commonly used 
(e.g. large-scale shear) this kind of forcing ensures the 
statistical isotropy and homogeneity of the vorticity field.}
\label{fig:1}
\end{figure}

\begin{figure}
\epsfxsize=240pt
\epsfbox{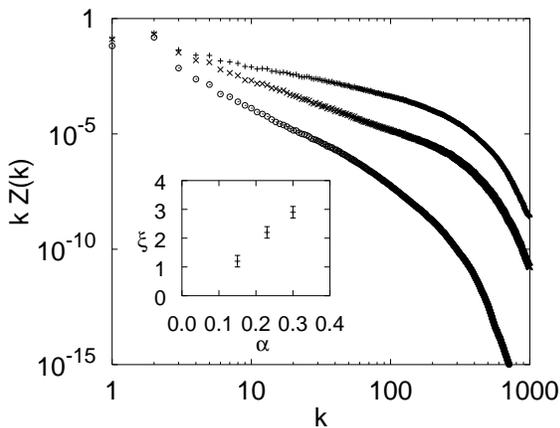}
\caption{The vorticity spectrum $Z(k) \sim k^{-1-\xi}$ steepens 
by increasing the Ekman coefficient $\alpha$. Here
$\alpha = 0.15$ $(+)$, 
$\alpha = 0.23$ $(\times)$,
$\alpha = 0.30$ $(\odot)$. In the inset, the exponent $\xi$ 
as a function of $\alpha$.}
\label{fig:2}
\end{figure}

\begin{figure}
\epsfxsize=240pt
\epsfbox{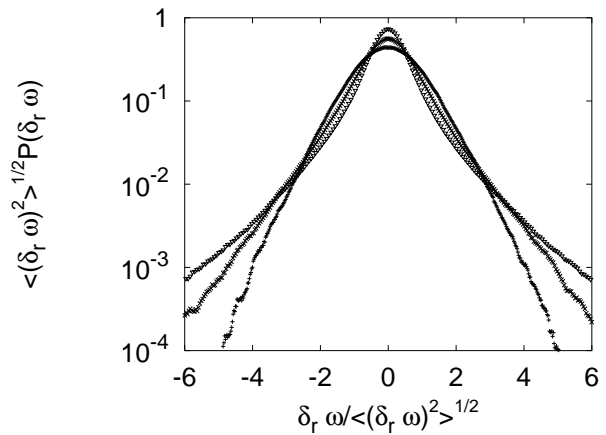} \vspace{0.2cm}
\caption{Probability density functions of normalized vorticity
increments $\delta_r \omega/\langle (\delta_r \omega)^2\rangle^{1/2}$.
Here, $r=0.20$ ($+$),  
$r=0.07$ ($\times$), $r=0.02$ ($\bigtriangledown$).
For large separations the statistics is close to Gaussian,
becoming increasingly intermittent for smaller $r$.}
\label{fig:3}
\end{figure}

\begin{multicols}{2}  
\narrowtext
In many physical situations, the incompressible flow of a shallow 
layer of fluid can be described by the two-dimensional Navier-Stokes
equations supplemented by a linear damping term which accounts
for friction. An important example comes from geophysical
rotating flows subject to Ekman friction \cite{nota}.
The dynamics can be written in terms of a single 
scalar field, the vorticity $\omega= {\bm \nabla} \times {\bm v}$, 
which obeys the equation  
\begin{equation}            
{\partial \omega \over \partial t} + {\bm v} \cdot \nabla \omega =
\nu \nabla^2 \omega - \alpha \omega + f_{\omega} \;,
\label{eq:1}
\end{equation}
where $\nu$ is the fluid viscosity, and $\alpha$ is the Ekman 
friction coefficient. 
The term $f_{\omega}$ is an external source of energy 
-- e.g. stirring --  that counteracts the dissipation
by viscosity and friction and allows to obtain a statistically
steady state. Here, we will study the statistical properties of
vorticity fluctuations $\delta_r \omega = \omega({\bm x}+{\bm r},t)-
\omega({\bm x},t)$ at scales $r$ smaller than the correlation length $L$
of the external forcing. We will show that the statistics of $\delta_r \omega$
is intermittent, and that the vorticity field
has the same scaling properties as a passive scalar with a finite lifetime. \\
As shown in Figure~\ref{fig:1}, the vorticity
field resulting from the numerical integration of Eq.~(\ref{eq:1})
is characterized by filamental structures whose thickness can be as
small as the smallest active lengthscales.
The wide range of scales involved in the vorticity dynamics manifests itself
in the appearance of power-law scaling for 
the spectrum of vorticity fluctuations
$Z(k)=2\pi k \langle |\hat{\omega}({\bm k})|^2 \rangle \sim k^{-1-\xi}$.
As already shown by Nam {\it et al} \cite{ott00}, 
the spectral slope $-1-\xi$ depends
on the intensity of the Ekman drag: for the
frictionless Navier-Stokes case ($\alpha=0$) we have $\xi=0$; 
a non-vanishing friction regularizes the flow depleting
the formation of small-size structures and results in a steeper spectrum
(see Fig.~\ref{fig:2}). In the range $0<\xi<2$ the exponent $\xi$
coincides with the scaling exponent $\zeta_2$ of the second-order moment
of vorticity fluctuations $S^\omega_2(r)=\langle (\delta_r \omega)^2 \rangle
\sim r^{\zeta_2}$. 
Let us now focus on a given value of $\alpha$.
In Figure~\ref{fig:3} we show the probability density functions of
vorticity fluctuations $\delta_r \omega$ at various $r$, 
rescaled by their rms value $\langle (\delta_r \omega)^2 \rangle^{1/2}$.
As the separation decreases, we observe that the probability of 
observing very weak or very intense vorticity excursions increases 
at the expense of fluctuations of average intensity. This phenomenon goes
under the name of intermittency. Its visual counterpart is the organization
of the field into ``quiescent'' areas (the patches,
where vorticity changes smoothly) and ``active'' regions (the filaments,
across which the vorticity experiences relatively strong excursions). 
The dynamical origin of this phenomenon can be understood as follows
(see also Ref.~\cite{ott00}). Let us first notice that, 
for any $\alpha$ strictly positive and as far as the 
statistical properties in the scaling range are concerned, we can disregard
the viscous term in Eq.~(\ref{eq:1}).
In other words, this system shows no dissipative anomaly,
due to the presence of friction \cite{che98}.
In the inviscid limit ($\nu=0$),
Eq.~(\ref{eq:1}) can be solved by the method of characteristics yielding
the expression
$\omega({\bm x},t)=
\int_{-\infty}^{t} f_\omega({\bm X}(s),s)\,\exp[-\alpha(t-s)]\,ds$, where 
${\bm X}(s)$ denotes the trajectory of a particle transported by the
flow, $\dot{\bm X}(s)={\bm v}({\bm X}(s),s)$, ending at 
${\bm X}(t)={\bm x}$. The uniqueness of the trajectory ${\bm X}(s)$
in the limit $\nu \to 0$ is ensured by the fact that 
the velocity field is Lipschitz-continuous, 
as it can be seen from the velocity spectrum 
$E(k)=Z(k)/k^2\sim k^{-3-\xi}$, always steeper than $k^{-3}$ 
(see Fig.~\ref{fig:2}). 
We remark that for $\xi>0$ the second-order velocity structure function
is dominated by the IR contribution of the spectrum and thus
trivially displays smooth scaling independently of the value of
$\xi$. This is not the case for odd order structure functions that,
in the absence of enstrophy dissipative anomaly, display anomalous
scaling at the leading order \cite{be00}. We have checked that
this is indeed the case in our simulations.
Vorticity differences are then associated to couples of particles 
$\omega({\bm x}',t)-\omega({\bm x},t)=
\int_{-\infty}^{t}  [f_\omega({\bm X}'(s),s)-f_\omega({\bm X}(s),s)]
\exp[-\alpha(t-s)] \;ds$. Inside the time integral, the difference between
the value of $f_\omega$ at ${\bm X}'$ and that at ${\bm X}$ is 
negligibly small as long as the two particles lie at a distance smaller
than $L$, the correlation length of the forcing; conversely, when 
the pair is at a distance larger than $L$, it  
approximates a Gaussian random variable $\Omega$. We then have  
$\delta_r \omega \sim \Omega \int_{-\infty}^{t-T_L(r)}
\exp[-\alpha(t-s)]\;ds \sim \Omega \exp[-\alpha T_L(r)]$, 
where $T_L(r)$ is the time that
a couple of particles at distance $r$ at time $t$ takes to reach a
separation $L$ (backward in time).
Large vorticity fluctuations arise
from couples of particles with relatively short exit-times 
$T_L(r) \ll \langle T_L(r) \rangle$, whereas
small fluctuations are associated to large ones.
Since the velocity field is smooth, two-dimensional and incompressible,
particles separate exponentially fast and their statistics can be 
described in terms of finite-time Lyapunov exponent $\gamma$.
For large times, the random variable 
$\gamma$ reaches a distribution $P(\gamma,t) \sim 
t^{1/2} \exp[-G(\gamma) t]$. The Cram\'er function $G(\gamma)$ 
is  concave, positive, with a quadratic minimum in $\lambda$ 
(the maximum Lyapunov exponent) $G(\lambda)=0$, and 
its shape far from the minimum depends on the details of 
the velocity statistics \cite{ott,BJPV98}. 
Finite-time Lyapunov exponent
and exit-times are related by the condition 
$L=r\exp[\gamma T_L(r)]$. That allows to obtain for $r\ll L$ 
the following estimate for moments of vorticity fluctuations
\begin{equation}
S_p^{\omega}(r) \sim \langle \Omega^p \rangle  \int d \gamma 
\left({r \over L} \right)^{p \alpha + G(\gamma) \over \gamma} 
\sim \left({r \over L} \right)^{\zeta^{\omega}_p}\;.
\label{eq:2}
\end{equation}
The scaling exponents are evaluated from Eq.~(\ref{eq:2})
by a steepest descent argument as 
$\zeta^{\omega}_p = \min_{\gamma} \left\{ 
p,[p \alpha + G(\gamma)]/\gamma ]\right\}\,$ .
Intermittency manifests itself in the nonlinear dependence
of the exponents $\zeta^{\omega}_p$ on the order $p$.
It has to be noticed that the active nature
of $\omega$ has been completely ignored in the above arguments:
the crucial hypothesis in the derivation of Eq.~(\ref{eq:2}) is that
the statistics of trajectories be
independent of the forcing $f_{\omega}$. This is quite a nontrivial 
assumption, since it is clear that forcing may affect large-scale
vorticity and thus influence velocity statistics, but it can
be justified by the following argument.
The random variable $\Omega$ arises from forcing contributions
along the trajectories at times $s<t-T_L(r)$, whereas the 
exit-time $T_L$ is clearly determined by the evolution of the strain
at times $t-T_L(r)<s<t$. Since the correlation time of the strain is 
$\alpha^{-1}$, for $T_{L}(r) \gg 1/\alpha$ we might expect that
$\Omega$ and $T_L(r)$ be statistically independent.
This condition can be translated in terms of the finite-time Lyapunov 
exponent as $r \ll L \exp(-\gamma/\alpha)$ and thus at
sufficiently small scales it is reasonable to consider $\omega$ 
as a passive field. We remark that, were the velocity field non-smooth,
the exit-times would be independent of $r$ in the limit $r \to 0$ 
and the above argument would not be relevant. Therefore, the smoothness of
the velocity field plays a central role in the equivalence of
vorticity and passive scalar statistics for this system.
\begin{figure}[hb]
\epsfxsize=250pt
\centerline{\hspace{3cm}\epsfbox{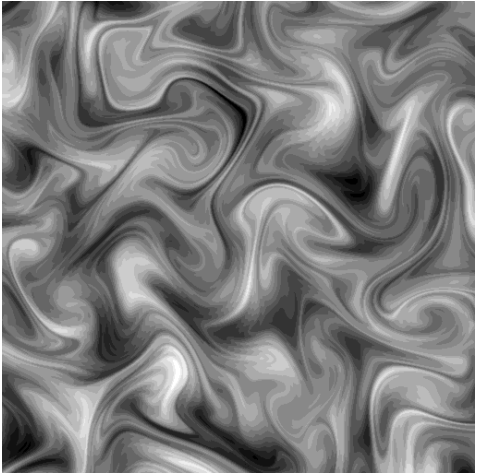}\vspace{0.2cm}}
\caption{Snapshot of the passive scalar field, simultaneous to the 
vorticity field shown in Fig.~\ref{fig:1}} 
\label{fig:4}
\end{figure}
To directly check whether small-scale vorticity can be considered
as passively advected by velocity, we also solved the equation
of transport of passive scalar with a finite lifetime \cite{che98,nam99,tel99}
\begin{equation}            
{\partial \theta \over \partial t} + {\bm v} \cdot {\bm \nabla} \theta =
\nu \nabla^2 \theta - \alpha \theta +  f_{\theta} \;.
\label{eq:3}
\end{equation}
where the velocity field results from the parallel integration of 
Eq.~(\ref{eq:1}). The parameters appearing in Eqs.~(\ref{eq:1})~
and~(\ref{eq:3}) are the same, yet the forcings $f_{\omega}$ and 
$f_{\theta}$ are independent processes with the same statistics.
According to the picture drawn above, 
we expect to observe the same small-scale 
statistics for $\delta_r \omega$ and 
$\delta_r \theta=\theta({\bm x}+{\bm r},t)-\theta({\bm x},t)$. 
\begin{figure}
\epsfxsize=240pt
\epsfbox{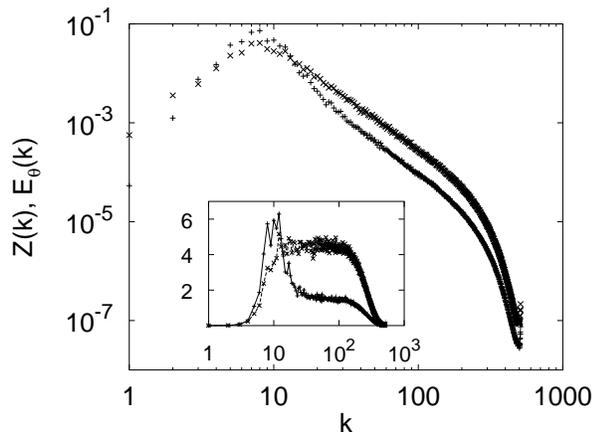}
\caption{Power spectra of  passive scalar ($\times$) and vorticity
($+$). Here $\alpha=0.15$.
In the inset we show the same spectra compensated by 
$k^{1+\zeta^{\theta}_2}$.} 
\label{fig:5}
\end{figure}
In Figure~\ref{fig:5} we show the power spectra of vorticity, $Z(k)$, and
of passive scalar $E_{\theta}(k)$. The two curves are parallel at large $k$,
in agreement with the expectation $\zeta^{\omega}_2=\zeta^{\theta}_2$.
We notice that the two spectra do not collapse exactly onto each other.
At large scales we observe a big bump in $Z(k)$ around $k=k_f$
which has not any correspondent in $E_{\theta}(k)$.
This deviation is due to the presence of an inverse energy flux 
in the Navier-Stokes equation, a phenomenon that has no equivalent 
in the passive scalar case. Due to this effect, the scaling quality
of $S^{\omega}_p(r)$ is poorer than the $S^{\theta}_p(r)$ one,
and a direct comparison of scaling exponents in physical space
is even more difficult.
\begin{figure}
\epsfxsize=240pt
\epsfbox{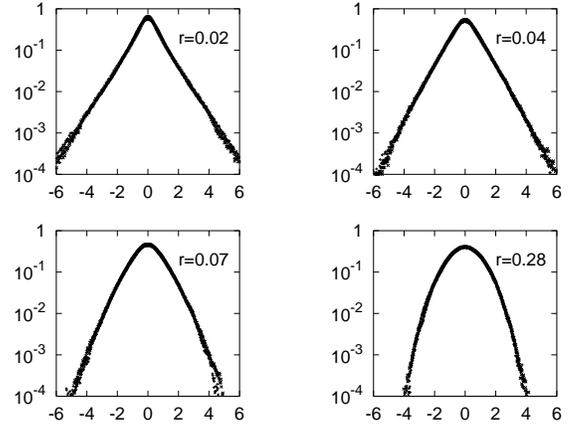}
\caption{Probability density functions of vorticity differences 
($+$) and of passive scalar ones ($\times$), 
normalized by their respective standard deviation,
at different scales $r$ within the scaling range.}
\label{fig:6}
\end{figure}
However, we observe in Fig.~\ref{fig:6} that the probability
density functions of vorticity and passive scalar increments,
once rescaled by their root-mean-square fluctuation, collapse 
remarkably well onto each other. 
That proves, along with the result
$\zeta^{\omega}_2=\zeta^{\theta}_2$ obtained from Fig~\ref{fig:5}, the
equality of scaling exponents of vorticity and passive scalar at any order:
$\zeta^{\omega}_p=\zeta^{\theta}_p$.

\begin{figure}
\epsfxsize=240pt
\epsfbox{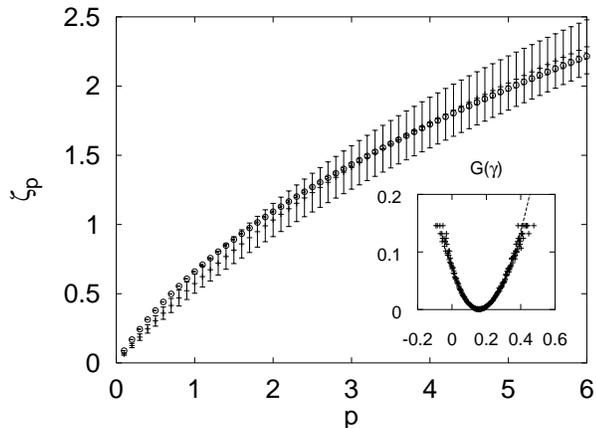}
\caption{The scaling exponents of the 
passive scalar $\zeta^{\theta}_p$ ($+$). 
We also show the exponents
obtained from the exit times statistics ($\odot$) according to
$\langle \exp[-\alpha p T_L(r)] \rangle \sim r^{\zeta_p^{\theta}}$ with 
average over about $2 \times 10^{5}$ couple of Lagrangian particles.
The errorbars are estimated by the r.m.s. fluctuation of the
local slope. In the inset we plot the Cramer function $G(\gamma)$
computed from finite time Lyapunov exponents (symbols) and exit
time statistics (line).}
\label{fig:7}
\end{figure}

The actual values can be directly extracted 
from the statistics of the passive scalar, which is not spoiled by 
large-scale objects. In Fig.~\ref{fig:7} we plot the first 
exponents $\zeta^{\theta}_p$ as obtained by looking at the local 
slopes of the structure functions $S_{p}^{\theta}(r)$.
The numerical values for $\zeta^{\theta}_p$ are
validated by the almost perfect agreement with the
Lagrangian exit-time statistics.

In conclusion, we have shown that in presence of linear friction
the small scale vorticity fluctuations in two dimensional
direct cascade are intermittent. Intermittency is 
a consequence of the the competition between exponential 
separation of Lagrangian trajectories and exponential
decay of fluctuations due to friction. 
Small-scale vorticity fluctuations behave statistically as a passive
scalar, as it has been
confirmed by a direct comparison. The smoothness of the velocity field
appears to be a crucial ingredient for the equality of active and 
passive scalar statistics. 

This work was supported by EU under the
contracts HPRN-CT-2000-00162 and FMRX-CT-98-0175.
Numerical simulations have been performed at IDRIS
(projects 011226 and 011411) and at CINECA (INFM Parallel Computing
Initiative).


\end{multicols} 
\end{document}